\documentclass[runningheads]{llncs}
\usepackage[inline]{enumitem}
\usepackage[hidelinks]{hyperref} 
\usepackage{hypcap} 
\usepackage{graphicx}
\usepackage{soul}

\usepackage[table,usenames,dvipsnames]{xcolor}
\begin{document}
\title{Reinforcement Learning-supported AB Testing of Business Process Improvements: An Industry Perspective}
\titlerunning{AB-BPM: An Industry Perspective}
%
\author{Aaron Friedrich Kurz\inst{1,2}\orcidID{0000-0002-2547-6780} \and
Timotheus Kampik\inst{2}\and
Luise Pufahl\inst{3} \and 
Ingo Weber\inst{3,4}\orcidID{0000-0002-4833-5921}}
\authorrunning{A. F. Kurz et al.}
%
\institute{Technische Universität Berlin, Berlin, Germany \\
\and SAP Signavio, Berlin, Germany\\
\email{\{aaron.kurz,timotheus.kampik\}@sap.com}
\and Technische Universität München, Munich, Germany\\
\email{\{luise.pufahl,ingo.weber\}@tum.de}
\and Fraunhofer Gesellschaft, Munich, Germany}
\maketitle              
\begin{abstract}
In order to better facilitate the need for continuous business process improvement, the application of DevOps principles has been proposed. In particular, the AB-BPM methodology applies AB testing and reinforcement learning to increase the speed and quality of improvement efforts. In this paper, we provide an industry perspective on this approach, assessing requirements, risks, opportunities, and more aspects of the AB-BPM methodology and supporting tools. Our qualitative analysis combines grounded theory with a Delphi study, including semi-structured interviews and multiple follow-up surveys with a panel of ten business process management experts. The main findings indicate a need for human control during reinforcement learning-driven experiments, the importance of aligning the methodology culturally and organizationally with the respective setting, and the necessity of an integrated process execution platform.

\end{abstract}
\section{Introduction}
Business processes are crucial for creating value and delivering products and services. 
Improving these processes is essential for gaining a competitive edge and enhancing value delivery, as well as increasing efficiency and customer satisfaction. 
This makes business process improvement (BPI) a key aspect of business process management (BPM), which is described as ``the art and science of overseeing how work is performed in an organization to ensure consistent outcomes and to take advantage of improvement opportunities''~\cite{dumas_fundamentals_2013}. 

DevOps, an integration of development and operations, is ``a set of practices intended to reduce the time between committing a change to a system and the change being placed into normal production, while ensuring high quality''~\cite{bass_devops_2015} and widely applied in the software industry.
A new line of research in the field of BPM proposes using DevOps principles, like AB testing, to facilitate continuous BPI with a method called \emph{AB-BPM}~\cite{Satyal:2019:IS}. 
AB testing assesses different software feature versions in the production environment with real users, usually in end user-facing parts of the software. The current version is only retired if the test data supports the improvement hypothesis. Applying such rapid validation of improvements to processes is a departure from the traditional BPM lifecycle, where the possibility of not meeting improvement hypotheses in production is rarely considered, leading to expensive do-overs~\cite{Satyal:2019:IS,holland_breakthrough_2005}.
Going beyond traditional AB testing, AB-BPM proposes the application of reinforcement learning (RL) to utilize performance measurements already while the experiments are conducted by dynamically routing incoming process instantiation requests to the more suitable version.

The AB-BPM method has not yet been systematically analyzed regarding the needs of BPM practitioners.
To facilitate applicability, additional research is needed to increase confidence in the proposed approach. Furthermore, BPM practitioners' insights on the AB-BPM methodology can be of value to the wider BPM community, since they can uncover hurdles and possibilities on the path to more automation in the field of process (re-)design. 
Therefore, this paper presents a qualitative study, with the overarching \textit{research question} being: What do BPM experts think about AB-BPM, and which implications does this have for the further development of the methodology and supporting tools?
To this end, we collected data on experts' views regarding the impact, advantages, and challenges of the AB-BPM method in an industry setting. In particular, we study the overall sentiment, perceived risks, potential use cases, technical feasibility, and software support requirements of the AB-BPM method.
To obtain the results, a panel of BPM experts from a large enterprise software company was interviewed and participated in follow-up surveys. We applied a mixture of the grounded theory~\cite{chun_tie_grounded_2019} and Delphi~\cite{dalkey_experimental_1963} research methodologies.

\section{Background and Related Work}\label{sec:backgr}
\begin{figure}[b]
\centering
\vspace{-1.5em}
\includegraphics[width=0.6\textwidth]{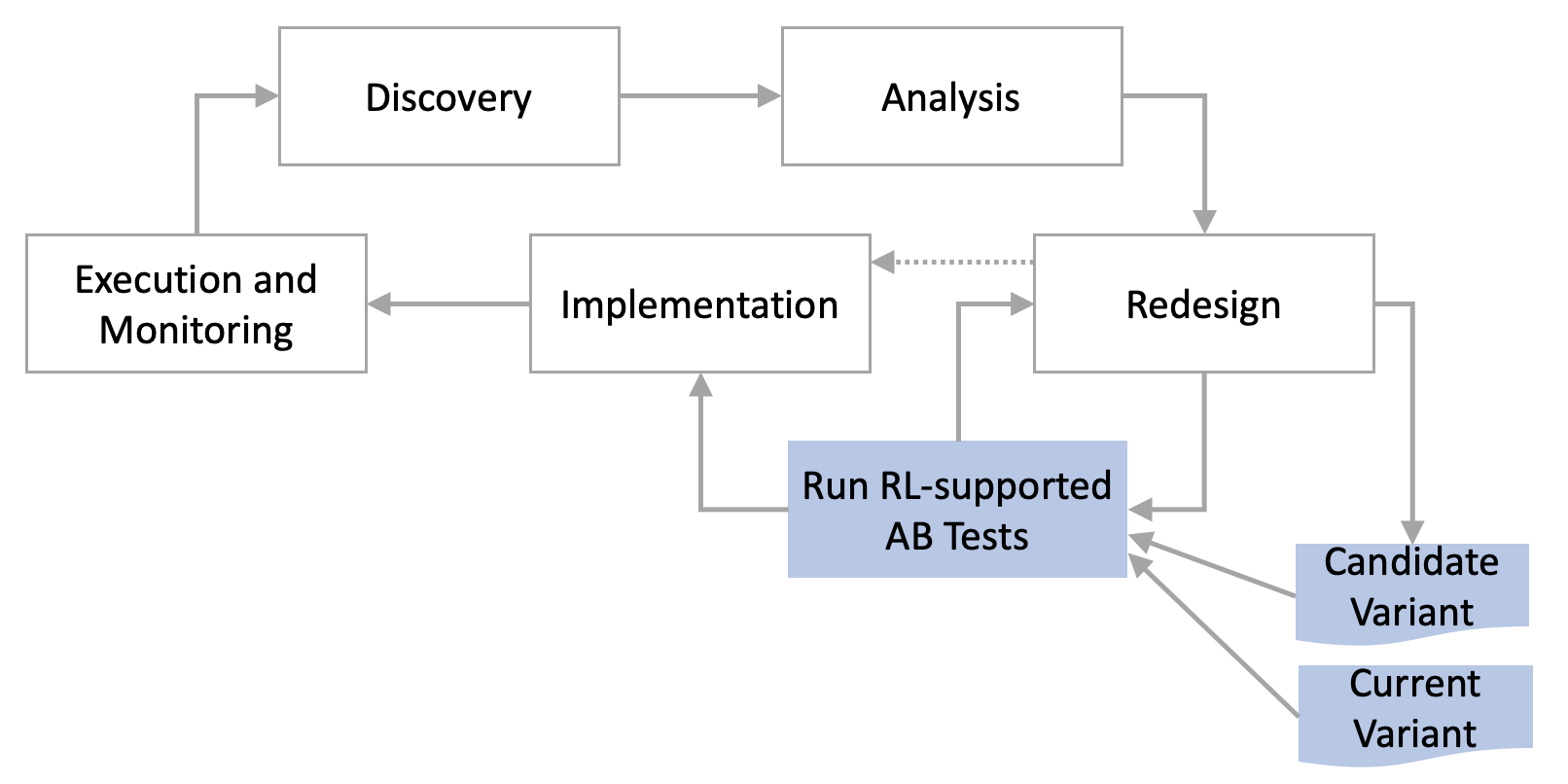}
\vspace{-1em}
\caption{Traditional BPM/AB-BPM lifecycle, adapted from \cite{Satyal:2019:IS,dumas_fundamentals_2013}; dotted arrow represents flow in traditional BPI lifecycle, blue areas represent additions in AB-BPM lifecycle.}
\label{fig:ab-bpm-bpm-lc}
\end{figure}

Organizations of all industries and sizes perform various combinations of activities to achieve desired results, may it be the production of physical goods or the provision of services. These combinations of activities are called business processes~\cite{aguilar-saven_business_2004}. They are often standardized and documented, meaning that each time an organization tries to achieve a particular result, they use a similar combination of activities. Such a standardized business process is often modeled graphically with the Business Process Model and Notation (BPMN)~\cite{von_rosing_business_2015}.
BPI is a central part of BPM~\cite{dumas_fundamentals_2013}, and also a core topic of this study. The traditional BPM lifecycle (see non-blue areas and dotted arrow of Figure \ref{fig:ab-bpm-bpm-lc}) is generally sequential~\cite{dumas_fundamentals_2013} and does not consider failures as ``first-class citizens''. This means that failures are not considered a necessary part of improvement and evaluated systematically, but rather a nuisance caused by insufficient planning in the redesign phase; and if failure happens, the whole lifecycle has to be repeated.

A more comprehensive approach to assessing the effects of change is AB testing.
The main goal of AB testing is to quickly determine whether a particular change to a system component will improve important performance metrics~\cite{bass_devops_2015}. The initial and the updated version of the component are tested in parallel using randomized experiments in a production environment (A vs B). The new version is often only made available to a select group of consumers, limiting any potential adverse impacts. The method has emerged as a popular approach when updating business-to-consumer software systems~\cite{sammut_online_2017}.

A particularly dynamic approach to AB testing can be facilitated by RL.
While supervised learning aims to learn how to label elements from a collection of labeled data, and unsupervised learning tries to find hidden structures in unlabeled data, RL has the goal of optimizing the behavior of a software agent based on a numeric reward in an interactive environment~\cite{sutton_reinforcement_2018}.
In RL, the agent is situated in a specific environment and has to decide on an action. This action is then evaluated, and a reward is calculated based on the action's consequence. The goal of the agent is to maximize the total reward it obtains. Learning which choices to make in what situation happens, essentially, through a systematic approach to trial and error~\cite{sutton_reinforcement_2018}.

As mentioned above, the sequential approach of the traditional BPI lifecycle fails to rapidly react to improvement hypotheses that are falsified in reality. This is a crucial shortcoming: research on BPI has shown that 75 percent of BPI ideas did not lead to an improvement: half of them had no effect, and a quarter even had detrimental outcomes~\cite{holland_breakthrough_2005}. This issue can be observed across domains: in a study conducted at Microsoft, only a third of the website improvement ideas actually had a positive impact~\cite{kohavi_controlled_2009}.
Furthermore, comparing process performance before and after the implementation is problematic in and of itself because changing environmental factors may be the primary driver of changes in process performance (or lack thereof).
To mitigate these problems, \cite{Satyal:2019:IS} proposes using AB testing when transitioning from the analysis to the implementation phase. This would mean that the redesigned version is deployed in parallel to the old process version, allowing for a fair comparison. Since  AB testing is not traditionally used in such a high-risk and long-running setting as BPM, the authors \cite{Satyal:2019:IS} apply RL to facilitate dynamic testing. With RL algorithms, we can make decisions based on the obtained data faster, by already dynamically routing process instantiation requests to the better-performing version during the experiment itself, thereby minimizing the risk of exposing customers to suboptimal process versions for too long. Altogether, AB-BPM should also allow for a shorter theoretical analysis of the redesign, in line with the DevOps mantra ``fail fast, fail cheap''. Figure~\ref{fig:ab-bpm-bpm-lc} (including the blue areas, excluding the dotted arrow) presents the improved AB-BPM lifecycle.

In addition to the RL-supported AB testing of improvement hypotheses, the complete AB-BPM method proposes some more test and analysis techniques. Our inquiry  focuses on the RL-supported AB testing of process variants: it is at the AB-BPM method's core, whereas the other steps in AB-BPM merely support the design of the RL-supported AB tests. References to AB-BPM in this work solely refer to the RL-supported AB testing of business process variants.

\section{Research Method}\label{sec:methodology}
In order to address the research questions, we apply the grounded theory (GT) research method~\cite{chun_tie_grounded_2019}, combined with the Delphi method~\cite{pare_systematic_2013}.
The GT research methodology is suitable for building a theory and answering questions about fields where little is known.
It has been selected because so far, no research has been conducted on the industry perspective of applying AB testing for BPI. After the initial data collection with semi-structured interviews (\textit{purposive sampling} in GT), we approached the second stage of the GT methodology (also called \textit{theoretical sampling}~\cite{chun_tie_grounded_2019}) as a shortened Delphi method study.
The novelty and complexity of the topic and the fact that the experts from the interviews have already been introduced to the AB-BPM method made a follow-up within this group of panelists more suitable than a broader follow-up survey, which would have required training the larger group and led to a more heterogeneous exposure of participant knowledge. The Delphi method has multiple sub-categories, and the version we use is called the ranking-type Delphi method (RTDM). The goal of RTDM is identifying and ranking key issues regarding a certain topic~\cite{pare_systematic_2013}. 
In the following paragraphs, we describe the research methods used in more detail.
\subsubsection*{Expert Selection.}
We have recruited experts from a multi-national software company with more than 100,000 employees. The company develops enterprise software, and the majority of study participants are employees of a sub-unit that specializes in developing BPM software. Due to the study's exploratory nature, the aim was to obtain a perspective from a broad range of experts. For this purpose, we set a number of goals for the selection of the experts: \emph{i)} to include people who develop BPM software as well as people that work in consulting (however, not necessarily both at once); \emph{ii)} to cover various areas of technical skills, e.g., software engineering and data science; \emph{iii)} the study participants should have experience with business process improvement initiatives.
The aim was to have a panel with ten experts, in line with standard practice for RTDM studies in Information Systems~\cite{pare_systematic_2013}. After reaching out to eleven people, ten people agreed to take part. Most study participants have a background in software engineering or other product-related roles (e.g., product management). But the panel also includes experts from consulting and a data scientist. The study participants had, on average, 7.6 years (SD = 2.4 years) of full-time industry experience, working in the BPM field for an average of 4.3 years (SD = 1.8 years).
Most of the experts (seven) have a degree in the Science/Engineering realm, while some (three) obtained their education in the field of Business/Management.
The highest educational degree of five of the study participants is a Ph.D., for four a master's and for one a bachelor's degree. The experts went through three study rounds: the interview, a validation survey and a ranking survey. Regarding participation levels, there was a drop from ten to five in the validation survey, whereas the final ranking survey reached eight people. Since the ranking survey is more important for the final results and included the option to give feedback on the coding as well, the level of participation after the initial interviews can be seen as relatively high.
\subsubsection*{Interviews.} As is common in GT research, we conducted semi-structured qualitative interviews with subject matter experts, aiming to capture a wide range of ideas and thoughts on the complexities of a topic by openly engaging in conversation with subjects and analyzing these interactions~\cite{robson_real_1999,brinkmann_interviews_2014}.
Since the order and wording of questions and the follow-up questions are highly flexible in semi-structured interviews, the interview guide is more of a collection of topics to be covered and not a verbatim script. There have also been minor adjustments to the interview guide during the interview phase in response to gained knowledge, in line with standard practice. Such adjustments are considered unproblematic since the goal is not a comparison of different subgroups, to test a hypothesis, or to find out how many people hold certain beliefs, but to find out what kind of beliefs are present~\cite{brinkmann_interviews_2014}. We used the following interview guideline, given in a condensed version:
\begin{enumerate*}
    \item prior experience with BPI,
    \item short introduction to AB-BPM (not a question, short presentation; 5-10 minutes),
    \item execution of AB tests/feasibility,
    \item suitability,
    \item prerequisites to adopt the AB-BPM method,
    \item risks,
    \item tool requirements,
    \item open discussion.
\end{enumerate*}
\subsubsection*{Consolidation and Validation.}
After the interviews, the transcripts were coded, and topics were consolidated (GT phase \textit{initial coding}~\cite{chun_tie_grounded_2019}). After the consolidation, the categories \emph{risks} and \emph{tool features} were selected for further data collection. The selection was motivated by the fact that the experts seemed highly interested in and provided many ideas around these categories; also, the categories can be considered highly relevant for the elicitation of requirements. The item lists were sent to the experts, which then had to validate whether their stance on the issues was properly represented. If not, the experts could give feedback on which items were missing or if some points should be separated and specified more clearly. Note that the narrowing-down phase, which asks the experts to exclude the least important items from each list, was skipped because the lists we presented to the experts had less than 20 items in them, to begin with. This is in accordance with common practice and guidelines~\cite{pare_systematic_2013,okoli_delphi_2004}.
\subsubsection*{Ranking.}
After validating the relevant points, the ranking phase aims to rank the items -- often with respect to the importance of issues. Since our two different lists, i.e., regarding risks and tool features, are topically distinct, we operationalized the ranking metrics differently for each list. Multiple rounds of ranking, as is common in RTDM studies, were outside of the scope of this work due to the extensive interviews and the focus on the exploration of new insights rather than the quantification of known facts.
Since \emph{risk} is a complex and hard-to-poll topic, we operationalized it as the product of the perceived likelihood of occurrence and the potential damage if said situation manifests~\cite{renn_concepts_2008}. The participants were asked to rank each the probability and the impact on a Likert scale: very low~(1) - low~(2) - moderate~(3) - high~(4) - very high~(5). This results in risk scores from 1 to 25.
Furthermore, we asked the study participants to rate the importance of possible \emph{tool features} on a Likert scale. The possible choices were: extremely unimportant~(1) - somewhat unimportant~(2) - neutral~(3) - somewhat important~(4) - extremely important~(5). 

\section{Results}\label{sec:results}
The main insights from the interviews and questionnaires are outlined below. 
We present opportunities and challenges that BPM industry experts perceive regarding the AB-BPM method. As software vendor employees, the study participants' answers, to some extent, reflect the experience of the wider industry, i.e., the customers' challenges. Statements of the experts are marked as quotations. 

\subsubsection*{Opportunities.}\label{sec:opportunities}
The \textbf{sentiment} towards AB-BPM was mainly positive. Multiple consultants brought up that some companies they worked with tried testing new process versions and comparing them with the status quo. However, the tests were mostly unstructured and considered only of a few instances or even no ``real'' instances (i.e., only tests). This means that any drawn conclusions are not dependable, due to the low number of instances and lack of statistical rigor when it comes to controlling confounding factors. Thus, AB testing provides a useful process improvement method that supports the structured testing of alternative versions. 

One question to the study participants was about the \textbf{suitability} of the method regarding contextual factors. The study participants were not presented with a list of categories but were free to elaborate on their intuition. More concretely, they were asked for what type of processes and what surrounding circumstances (company, market, industry) they think the methodology would be well- or ill-suited. Their statements were then mapped to the categorization of BPM contexts by~\cite{brocke_role_2016}. The result can be seen in Table~\ref{tab:usecases}. The characteristics in italics present special cases for factors where every characteristic was deemed suitable, which we will outline in the following.

\begin{table}[!ht]
    \centering
    \addtolength{\leftskip} {-4cm} 
    \addtolength{\rightskip}{-4cm}
    \scalebox{0.85}{
    \begin{tabular}{|l|lll|}
    \hline
        \textbf{Contextual factors} & \textbf{Characteristics} & \textbf{} & \textbf{} \\ \hline
        \textbf{Goal-dimension} & ~ & ~ & ~ \\ 
        Focus & \cellcolor{LimeGreen!40}\textit{Exploitation (Improvm., Compl.)} & \cellcolor{LimeGreen!40}Exploration (Innovation) & ~ \\ 
        \textbf{Process-dimension} & ~ & ~ & ~ \\ 
        Value contribution & \cellcolor{LimeGreen!40}\textit{Core process} & \cellcolor{LimeGreen!40}Management process & \cellcolor{LimeGreen!40}Support process \\ 
        Repetitiveness & \cellcolor{LimeGreen!40}Repetitive & \cellcolor{Orange!40}Non-repetitive & ~ \\ 
        Knowledge-intensity & \cellcolor{LimeGreen!40}Low knowledge-intensity & \cellcolor{Goldenrod!40}Medium knowledge-intensity & \cellcolor{Orange!40}High knowledge-intensity \\ 
        Interdependence & \cellcolor{LimeGreen!40}Low interdependence & \cellcolor{Goldenrod!40}Medium interdependence & \cellcolor{Orange!40}High interdependence \\ 
        Variability & \cellcolor{LimeGreen!40}Low variability & \cellcolor{Goldenrod!40}Medium variability & \cellcolor{Orange!40}High variability \\ 
        \textbf{Organization-dimension} & ~ & ~ & ~ \\ 
        Scope & \cellcolor{LimeGreen!40}Intra-organizational process & \cellcolor{Orange!40}Inter-organizational process & ~ \\ 
        Industry & \cellcolor{Orange!40}Product industry & \cellcolor{Goldenrod!40}Product \& service industry & \cellcolor{LimeGreen!40}Service industry \\ 
        Size & \cellcolor{Orange!40}Start-up & S\cellcolor{Goldenrod!40}mall and medium enterprise & \cellcolor{LimeGreen!40}Large organization \\ 
        Culture & \cellcolor{LimeGreen!40}Highly supportive of BPM & \cellcolor{Goldenrod!40}Medium supportive of BPM & \cellcolor{Orange!40}Non-supportive of BPM \\ 
        Resources & \cellcolor{Orange!40}Low organizational resources & \cellcolor{Goldenrod!40}Medium organizational resources & \cellcolor{LimeGreen!40}High organizational resources \\ 
        \textbf{Environmental-dimension} & ~ & ~ & ~ \\ 
        Competitiveness & \cellcolor{LimeGreen!40}Low competitive & \cellcolor{LimeGreen!40}Medium competitive & \cellcolor{LimeGreen!40}\textit{Highly competitive} \\ 
        Uncertainty & \cellcolor{LimeGreen!40}Low env. uncertainty & \cellcolor{Goldenrod!40}Medium env. uncertainty & \cellcolor{Orange!40}High env. uncertainty \\ \hline
    \end{tabular}}
    \vspace{0.25cm}
    \caption{Suitability of AB-BPM method regarding BPM context, color coding: \colorbox{LimeGreen!40}{green - high suitability}, \colorbox{Goldenrod!40}{yellow - medium suitability}, \colorbox{Orange!40}{orange - low suitability}. Categorization from \cite{brocke_role_2016}. Items in italics present particularly interesting/suitable cases for factors where every characteristic is suitable.}
    \label{tab:usecases}
    \vspace{-0.75cm}
\end{table}

\begin{description}
\item{\textbf{Focus.}} No agreement could be reached on whether AB-BPM was suitable for radical changes. \cite{Satyal:2019:IS} present the method primarily for evolutionary changes, while some study participants believe it is suitable for both. However, most consider the method more appropriate for small process changes due to the ease and speed of implementation. AB-BPM would be incompatible with fundamental changes that require ``lengthy discussions'' and expensive financial obligations, making rapid testing difficult. Somewhat larger changes within the same information system may be feasible, but smaller changes are generally preferred.
\item{\textbf{Value contribution.}} Using the AB-BPM method might be especially useful in core processes. This is because other processes are found at many companies and cannot be used for meaningful differentiation. One study participant noted that it is advisable to ``differentiate where you differ.'' They said, ``as a sports shoe company, we could strive to have the best finance processes, but that won't make people buy our shoes -- we need better shoes and better shoe quality to win in the marketplace.'' They, therefore, recommended using standard processes for everything but the core processes. This is already common practice and also suggested by academic studies~\cite{lubke_effectively_2019}. For core processes, however, experimentation with the AB-BPM method would be highly favorable.
\item{\textbf{Competitiveness.}} In general, there were no opinions indicating that any level of market competitiveness would lead to less suitability of the method. Study participants noted, however, that highly competitive markets would increase the need for such a tool to allow for faster process iterations, ``to stay competitive.''
\end{description}

The elicitation of \textbf{requirements} for a tool that executes and supports the AB-BPM method was also part of the study, and the identified feature requirements are presented in the following. First, we present the ranking of the items (see Table~\ref{tab:rank_towi}). Afterward, more details on the items ranked as most important are provided.

The ranking is based on the importance Likert scale presented in Section \ref{sec:methodology}. The average importance scores (AVG) are accompanied by the standard deviations (SD) to give an insight into the level of agreement among the experts. Furthermore, the feature requirements have been categorized into presentation, procedure, and support. This categorization has been created after and based on the interviews, during the coding of the interviews (GT phase \textit{intermediate coding} \cite{chun_tie_grounded_2019}).
 \emph{Presentation} includes features regarding the presentation of data, or features that are more focused on the front end of the tool in general;
\emph{Procedure} are features regarding the underlying technical or methodological procedure;
\emph{Support} includes features that already exist in the AB-BPM method but that have not been presented to the study participants during the introduction to AB-BPM (see Section \ref{sec:backgr}); they, therefore, support the equivalent suggestions by~\cite{Satyal:2019:IS}.

\begin{table}[!ht]
    \centering
    \addtolength{\leftskip} {-2cm} 
    \addtolength{\rightskip}{-2cm}
    \scalebox{0.85}{
    \begin{tabular}{|l|l|l|l|}
    \hline
        \textbf{Code} & \textbf{Tool Feature} & \textbf{Imp. AVG} & \textbf{Imp. SD} \\ \hline
        \cellcolor{Salmon!35}COM & Communicating process changes efficiently for teaching and enablement of employees & 4.75 & 0.46 \\ \hline
        \cellcolor{ProcessBlue!35}DIF & BPMN diff viewer & 4.57 & 0.53 \\ \hline
        \cellcolor{Salmon!35}PIB & See potential impact beforehand (amount and business-wise) & 4.43 & 0.79 \\ \hline
        \cellcolor{Salmon!35}ETL & Exec. on/with various systems; ETL from various systems for data extraction & 4.29 & 0.95 \\ \hline
        \cellcolor{ProcessBlue!35}IAR & Clear insights and action recommendations & 4.25 & 0.71 \\ \hline
        \cellcolor{Salmon!35}EES & Emergency exit/stop & 4.25 & 1.04 \\ \hline
        \cellcolor{Salmon!35}CPS & Capture process participants sentiments and feedback on process variants & 4.13 & 1.13 \\ \hline
        \cellcolor{ProcessBlue!35}DDI & Detailed/drill-down insights & 4.13 & 0.64 \\ \hline
        \cellcolor{gray!35}IWS & Integrate with simulation & 4.13 & 0.83 \\ \hline
        \cellcolor{Salmon!35}BRK & Offer broad range of possible KPIs to take into account & 4.00 & 0.93 \\ \hline
        \cellcolor{ProcessBlue!35}EMA & Show analytics embedded in process diagram & 4.00 & 0.58 \\ \hline
        \cellcolor{gray!35}HID & Use of historical data & 4.00 & 0.58 \\ \hline
        \cellcolor{Salmon!35}PSN & Pre-setting stop and notification criteria & 3.88 & 0.83 \\ \hline
        \cellcolor{Salmon!35}EXC & Potential exclusion of certain customer groups & 3.86 & 1.07 \\ \hline
        \cellcolor{gray!35}REC & Randomization/not always choose same employees for test & 3.50 & 0.76 \\ \hline
        \cellcolor{Salmon!35}VRC & Result can be different variants for recognized customer patterns & 3.50 & 0.84 \\ \hline
        \cellcolor{Salmon!35}MRL & Options for modification of reinforcement learning-based routing & 3.43 & 0.98 \\ \hline
        \cellcolor{Salmon!35}XRL & Offer explainable reinforcement learning & 3.25 & 1.28 \\ \hline
        \cellcolor{Salmon!35}MTT & Experiment with more than two variants & 3.13 & 1.46 \\ \hline
    \end{tabular}}
    \vspace{0.25cm}
    \caption{Item list of desired tool features, in order of perceived importance. Colors in ``Code'' column depict categories: \colorbox{ProcessBlue!35}{blue - presentation}, \colorbox{Salmon!35}{pink - procedure}, \colorbox{gray!35}{gray - support}.}
    \label{tab:rank_towi}
    \vspace{-0.75cm}
\end{table}

In the following, the three tool features ranked as most important are described in more detail.
\begin{description}
\item{\textbf{See potential impact beforehand (amount and business-wise).}}
According to the study participants, process experts should be able to see estimations on possible impacts beforehand to support an informed decision-making process, e.g., how many customers or what order volume would be affected by the test.
\item{\textbf{BPMN diff viewer.}}
One study participant emphasized the importance of human experts having a clear understanding of changes in the current initiative. They suggested a "diff viewer for the diagrams" - a graphical representation of changes made to a document from one version to another, commonly used in software engineering. In business process management, this could involve versions of a BPMN diagram, with changes highlighted in different colors. Diff viewers are well-researched and applied in this context, for example in~\cite{pietsch_comparison_2012,ivanov_bpmndiffviz_2015}.
\item{\textbf{Communicating process changes efficiently for teaching and enablement of employees.}}
The need for process participants to learn how new versions have to be executed was stressed by multiple interviewees. One study participant stated that ``one needs to notify the people working on steps in the process of the changes.'' More ``enablement is needed to teach employees the changes,'' and another study participant noted that ``seeing how this [aspect of change management] can be integrated would be an interesting question.'' This would go beyond just teaching single steps but also create openness and transparency about goals and project setup, allowing for ``a lot of change, even in parallel, without people being lost.''
Similar to the diff viewer, change notification management is a feature that has already received research attention in the context of business process management software~\cite{yan_business_2012,rosa_apromore_nodate}.
\end{description}

\subsubsection*{Challenges.}\label{sec:challenges}
It is vital to know the core challenges to advance the AB-BPM methodology and adjacent endeavors. Only then can they be addressed and mitigated adequately. The risks and further challenges that the expert panel has voiced are, therefore, outlined below.

A critical goal of this work is to determine the AB-BPM method's principal \textbf{risks} since they hinder its usage and implementation in organizations.
In the following, we will present the results from the experts' ranking and then give a more detailed outline of the most highly ranked individual risks.
The risks, alongside their AVG risk scores and the SD of those scores, can be seen in Table~\ref{tab:rank_risk}. Furthermore, the risks have been categorized as follows.
\emph{Culture} are risks regarding the working culture and employees of the company. \emph{Results} include risks regarding results, decisions, and outcomes; \emph{Operations} consists of risks regarding the implementation and execution of the AB-BPM method itself, but also the normal business operations; \emph{Legal} includes risks regarding the cost and loss of income caused by legal uncertainty~\cite{tsui_experience_2013}.

\begin{table}[!ht]
    \centering
     \addtolength{\leftskip} {-2cm} 
    \addtolength{\rightskip}{-2cm}
    \scalebox{0.85}{
    \begin{tabular}{|l|l|l|l|}
    \hline
        \textbf{Code} & \textbf{Risk} & \textbf{Risk AVG} & \textbf{Risk SD} \\ \hline
        \cellcolor{ProcessBlue!35}CHM & Change management problems during rollout of new variants (cultural) & 15.6 & 4.8 \\ \hline
        \cellcolor{Salmon!35}BLE & Blindly following machine-generated analysis results leading to erroneous decisions & 15.5 & 5.7 \\ \hline
        \cellcolor{Salmon!35}UVD & Unclear results due to high process variance and process drift & 14.4 & 5.4 \\ \hline
        \cellcolor{Salmon!35}IGK & Problems due to improperly set goals/KPIs & 14.3 & 7.0 \\ \hline
        \cellcolor{Salmon!35}ESU & Economically or societally undesirable bias & 13.6 & 7.5 \\ \hline
        \cellcolor{ProcessBlue!35}TST & Process participants (employees) feeling like test subjects/tracked & 11.3 & 5.8 \\ \hline
        \cellcolor{ProcessBlue!35}EPG & Employees purposely acting with a certain goal for the experiment in mind & 11.1 & 4.6 \\ \hline
        \cellcolor{LimeGreen!35}DNO & Disturbance of normal operations & 11.1 & 4.7 \\ \hline
        \cellcolor{LimeGreen!35}DIE & Problems with deploym./impl. and exec. of multiple process variants at same time & 11.0 & 5.6 \\ \hline
        \cellcolor{LimeGreen!35}SEP & Scaling and edge case problems & 10.4 & 6.8 \\ \hline
        \cellcolor{LimeGreen!35}FPI & Failed experimental process instances & 10.3 & 3.5 \\ \hline
        \cellcolor{LimeGreen!35}LRE & Lacking responsibility in case of problems & 10.0 & 8.2 \\ \hline
        \cellcolor{ProcessBlue!35}EDA & Employee dissatisfaction due to feeling like one is about to be automated & 10.0 & 3.8 \\ \hline
        \cellcolor{Salmon!35}RED & Reputational damage for process provider & 9.6 & 3.1 \\ \hline
        \cellcolor{gray!35}LCH & Legal challenges due to experiments & 9.1 & 5.4 \\ \hline
    \end{tabular}}
    \vspace{0.25cm}
    \caption{Item list of risks, in order of perceived risk. Colors in ``Code'' column describe categories: \colorbox{Salmon!35}{pink - results}, \colorbox{ProcessBlue!35}{blue - culture}, \colorbox{LimeGreen!35}{green - operations},
    \colorbox{gray!35}{gray - legal}.}
    \label{tab:rank_risk}
    \vspace{-0.75cm}
\end{table}
\vspace*{0.5cm}
In the following, details on the three most highly ranked risks are explained in more detail.
\begin{description}
    \item[\textbf{Unclear results due to high process variance and process drift.}] As mentioned before, the execution of business processes can differ from how they were intended to be executed and it is subject to (unintended) changes over time. This phenomenon, called process drift~\cite{sato_survey_2021}, leads to a high variance of executed process versions. This could pose a risk for the AB-BPM method since it is then unclear whether process participants execute the two versions as they are intended. A process participant is a company-internal actor performing tasks of a process~\cite{dumas_fundamentals_2013}, i.e., an employee of the organization executing the process. If the process cases vary from the intended way of execution, it is hard to draw conclusions from the results since they might be based on a change that occurred spontaneously instead of the planned process changes. One example might be that ``people exchange emails instead of following the steps in the process execution software.''
    \item[\textbf{Erroneous machine-generated analysis results are blindly followed.}] \phantom{e}\\ Many interviewees noted that solely relying on the algorithm's interpretation of the data might cause problems. One study participant noted that ``such models are always an abstraction of reality [...] and relying on them completely can lead to mistakes.'' This topic also came up during the discussion of bad prior experiences, when a study participant noted that sometimes wrong decisions were made because of a lack of understanding of data. One potential example is the use of team performance metrics, which are often highly subjective (e.g., workload estimates in some project management methods), without context. Putting data into context and not blindly following statistical calculations is, therefore, a core challenge that needs to be addressed.
    \item[\textbf{Cultural change management problems during variant roll-out.}] \phantom{e}\\ This risk was added after the validation survey since one study participant remarked that this item was missing. It can be understood as incorporating any other cultural change management issues not yet included in the item list (see blue items in Table~\ref{tab:rank_risk}). The high rating of this item can be seen as an indicator that the human side of the method and adjacent tools must not be left out of research and development efforts. The importance of culture also became very apparent when asked about \textit{prerequisites for the use of the AB-BPM method}. Many study participants noted that the organization would need to have an experiment culture, meaning that they should be open to trying new things and handling failures as learning opportunities. Furthermore, they stated a need for organizational transparency and trust.
\end{description}

The implementation and adoption of AB-BPM as presented in \cite{Satyal:2019:IS} assumes the existence of a Business Process Management System (BPMS) that allows for the direct deployment of BPMN models. A BPMS is an information system that uses an explicit description of an executable process model in the form of a BPMN model to execute business processes and manage relevant resources and data. It presents a centralized, model-driven way of business process execution and intelligence~\cite{dumas_fundamentals_2013}. However, most processes are executed by non-BPMS software, i.e., they are not executed from models directly~\cite{10.1007/978-3-031-07475-2_9}.

Therefore, whether the usage of a BPMS is a requirement for \textbf{technical feasibility} is a research question of this study. Altogether, AB testing of business processes seems technologically feasible without a BPMS, one interviewee noted: ``I do not think that it is a problem that processes are executed over several IT systems since you only need to be able to start either process version. The route they are going afterward, even if it is ten more systems, is no longer relevant.'' However, if we want to use live analytics to route incoming process instantiation requests (e.g., as proposed with RL) without a BPMS, we would need an Extract-Transform-Load (ETL) tool. ETL software is responsible for retrieving relevant data from various sources while bringing the data into a suitable format~\cite{vassiliadis2009survey}. Relying on a BPMS would not only have the benefit of easier data collection and access, it would also make deploying and executing experimental processes more straightforward. Furthermore, such an ETL tool might also be highly complex due to the many systems processes can potentially touch. One study participant noted that when a BPMS does not exist, ``you will have to put a lot of effort into mining performance data; it would be more difficult to get the same data from process mining, covering every path and system.'' In fact, most study participants did deem a BPMS, or something similar, to be a prerequisite. One study participant stated, however, that while some ``central execution platform'' would probably be needed, it remains unclear whether these have to be in the shape of current BPMS. Overall, there seemed to be the notion that the integrated, model-driven way of orchestrating and executing business processes offered by BPMS is the direction the industry should move towards. 

Besides the risks, the study participants also mentioned \textbf{other challenges}.
Here, we highlight some of them.
Regarding the question of \textit{bad prior experiences when conducting BPI initiatives}, some study participants criticized the unclear impact of process improvements during/after BPI projects. This is due to constantly changing environmental factors and the resulting difficulty to compare process data that has been collected at different points in time. This highlights the possible positive impact that AB-BPM could have on BPI efforts, by giving BPM experts a better data basis to evaluate improvement efforts.
Regarding the \textit{prerequisites for the use of the AB-BPM method}, on the more technical side, the interviewees noted that many companies would not offer the level of continuous data metrics needed for the dynamic, RL-driven routing during the experiments. This again highlights the need for an integrated process execution (e.g., a BPMS).

\section{Discussion}\label{sec:discussion}

Emerging patterns in the coded items can be brought together to form tool and methodological suggestions. This analysis of patterns in the coded interview data can be seen as an application of the GT phase \textit{advanced coding}~\cite{chun_tie_grounded_2019}. In the following, we present the inferred suggestions and concepts in more detail. These serve as both concrete suggestions and examples of how the requirements can be used when implementing AB-BPM tools.

\begin{description}
\item{\textbf{Integrated Process Model Repository.}}
The idea of integrating the AB-BPM tool with a process model repository (PMR) came up in one interview. It is implicitly supported by statements of most of the other study participants.
The main aim of IPMR is to provide a fast and efficient way of introducing new process variants to process participants while allowing them to provide feedback. Once a new business process variant is added, the responsible process expert would need to add additional material to help process participants understand the new version. Suppose a specific process participant is part of an experimental process case (or a new version is permanently rolled out). In that case, they would receive a notification (e.g., via email) with the most important information regarding the process update, with a link to more material. This could be extended with the need for process participants to pass a short test to avoid improvement effort failures due to misunderstanding. They could also provide feedback, which could then be evaluated by the process experts or even considered for the RL experiment.

\item{\textbf{Instance Recovery Mechanism.}}
The analysis of the interviews made it clear that the danger of failed process instances has to be mitigated in some way. Based on that data, we propose an instance recovery mechanism (IRM). Before starting an experiment, process experts could set thresholds for specific KPIs. The expert would be notified when a process instance reaches such a threshold. They could then intervene in the process execution to make sure that the process consumers still get their desired product or service. The thresholds should be set relatively high, to only intervene in erroneous cases. Otherwise, it could create a bias in the experiment. This is why one also has to consider how such instances are evaluated in calculating the rewards of the RL. Leaving them out of the RL calculations could lead to problems: Imagine only one process version sometimes exceeds the thresholds and has to be salvaged manually. By leaving these instances out, we might misjudge that version. A better approach could include these manually salvaged versions in the model with a certain penalty value.
Considering the scientific state-of-the-art, IRM is closely related to the notion of (real-time) business process monitoring, as for example covered in~\cite{pedrinaci_sentinel_2008,kang_real-time_2012}.
\end{description}

Another pattern that can be observed throughout the interviews is the need for more human control. This can be seen in the perceived risks (e.g., BLE, LRE), as well as in the desired tool features (e.g., EES, CPS, PSN, MRL, XRL). In a previous paper, we have already presented an initial prototype that enhances the AB-BPM methodology with additional features for human control \cite{kurz_hitl-ab-bpm_2022}. 

The presented study has several \textbf{limitations}.
One possible threat to the validity of results, especially the ranking, is that we employed a shortened RTDM. Usually, there are multiple ranking rounds until the concordance among the experts has reached a satisfactory level. As mentioned above, we decided not to conduct multiple ranking rounds due to the exploratory nature of the study and the extensive interview process, which makes it difficult to be certain in the ranking. Preliminary statistical analysis of the rating differences between the items shows that there is no significant difference between the listed risk items, but there is a significant difference between the ratings of the different feature items. However, given the qualitative nature of the study the rankings should only be seen as rough guidance. The focus of the work is the qualitative elicitation of items. This has been taken into consideration for the discussion and conclusion, meaning that we took all the items into account, not just the most highly ranked.

Another possible threat to the validity of the results is that the participation in the validation round dropped from ten to five. We tried to mitigate validity issues by ensuring higher participation in the ranking round (eight experts participated) and giving study participants the option to also give feedback on the coding in that round. Since no more remarks were made about the coding, we conclude that the coding was satisfactory for all eight study participants of the ranking round.

Furthermore, all study participants are employed by the same company. We tried to attenuate this by selecting experts with extensive experience from various teams and backgrounds. Additionally, the consultants brought in their experience with business process improvement projects from additional companies.

\section{Conclusion}\label{sec:conclusion}
The main aim of this study was to obtain practitioners' perspectives on the AB-BPM method. Using mostly qualitative research methods, we shed light on the requirements for the further development of AB-BPM tools and the underlying method.
Overall, the study participants perceived the methodology as advantageous in comparison to the status quo. The three main conclusions of the study are:
\emph{i)} more possibilities of human intervention, and interaction between the RL agent and the human expert, are a core requirement,;
\emph{ii)} transparency and features for the participation of process participants are needed to make AB-BPM culturally viable;
\emph{iii)} integrated process execution is necessary to facilitate the seamless deployment of parallel process variants and deliver the real-time data needed for dynamic RL and routing.
The openness of the semi-structured interviews facilitated the discovery of future research opportunities, e.g., studying companies carrying out unstructured process tests in a production environment, and tool-driven training of process participants. While we focused on a single process improvement method, practitioners' insights on different kinds of business improvement methods~\cite{malinova2022study} could provide insights for studies that compare AB-BPM to other methodologies.
%
%
%
\bibliographystyle{splncs04}
\bibliography{paper}
\end{document}